\documentclass{article}
\usepackage{graphicx}
\usepackage{amsmath}
\usepackage{amsfonts}

\usepackage{amssymb}
\let\UnmodifSec=\section
\renewcommand{\section}{\setcounter{equation}{0}\UnmodifSec}

\def\dateline{\today}
\newtheorem{definition}{Definition}[section]
\newtheorem{lemma}{Lemma}[section]

\newtheorem{theorem}{Theorem}[section]
\newtheorem{remark}{Remark}[section]
\newtheorem{corollary}{Corollary}[section]

\def\beq{\begin{equation}}
\def\endq{\end{equation}}
\def\beqa{\begin{eqnarray}}
\def\endqa{\end{eqnarray}}

\def\RR{\mathbb R}

\def\bC{{\bf C}}
\def\bR{{\bf R}}

\def\AA{{\cal A}}

\def\CC{{\cal C}}

\def\LL{{\cal L}}
\def\MM{{\cal M}}

\def\PP{{\cal P}}

\def\RR{{\cal R}}
\def\SS{{\cal S}}
\def\TT{{\cal T}}
\def\UU{{\cal U}}
\def\VV{{\cal V}}

\def\XX{{\cal X}}

\def\wh{\widehat}
\def\wt{\widetilde}
\def\ovl{\overline}

\def\HB{\hfill\break}
\def\interior#1{\setbox1=\hbox{$#1$}\rlap{$#1$}\kern0.4\wd1\raise1.1\ht1%
\hbox{$\scriptstyle \circ$}}
\def\bydef{\mathrel{\buildrel \hbox{\scriptsize \rm def} \over =}}
\def\boxit#1#2{\setbox1=\hbox{\kern#1{#2}\kern#1}%
\dimen1=\ht1 \advance \dimen1 by #1 \dimen2=\dp1 \advance \dimen2 by #1
\setbox1=\hbox{\vrule height\dimen1 depth\dimen2\box1\vrule}%
\setbox1=\vbox{\hrule\box1\hrule}%
\advance \dimen1 by .4pt \ht1=\dimen1 \advance \dimen2 by .4pt \dp1=\dimen2
\box1\relax}
\def\endprf{\raise .5ex\hbox{\boxit{2pt}{\ }}}

\def \rge{\lower 0.55ex\hbox{$\,\buildrel > \over {\sim}\,$}}
\def \rle{\lower 0.55ex\hbox{$\,\buildrel < \over {\sim}\,$}}

\def\SA{{(\SS||1,\ A)}}
\def\SPA{{(\SS'||1,\ A)}}
\def\SB{{(\SS||1,\ B)}}
\def\SPB{{(\SS'||1,\ B)}}

\def\DTA{{(1 \downarrow \TT ||1,\ A)}}
\def\UTPA{{(1 \uparrow \TT' ||1,\ A)}}
\def\DTB{{(1 \downarrow \TT ||1,\ B)}}
\def\UTPB{{(1 \uparrow \TT' ||1,\ B)}}
\def\UUVPA{{(1 \uparrow 2 \uparrow \VV'||1,\ A)}}
\def\DDVA{{(1 \downarrow 2 \downarrow \VV||1,\ A)}}
\def\UVPC{{(1\uparrow \VV'||1,\ C)}}
\def\DVC{{(1\downarrow \VV||1,\ C)}}
\def\ifundefined#1{\expandafter\ifx\csname#1\endcsname\relax}
\def\lineto{\smash{\hbox{$\relbar\kern -0.24em\relbar$}}}
\def\arto{\smash{\hbox{$\rightarrow\kern -0.24em\relbar$}}}
\def\arfr{\smash{\hbox{$\relbar\kern -0.24em\leftarrow$}}}
\def\bu#1{\hbox{\kern -0.2em$\buildrel{#1}\over\bullet$\kern -0.2em}}

\def\vertup{\kern-2.3ex}
\def\colbu#1#2#3{\hbox{\raise -5ex \vbox{\normalbaselineskip=3.46 ex 
\baselineskip=3.46 ex 
\nointerlineskip%
\hbox{\kern -0.21em$\buildrel{#1}\over\bullet$\kern -0.2em}
\vertup\hbox{\kern-0.1em$|$}%
\vertup\hbox{\kern-0.1em$|$}%
\vertup\hbox{\kern -0.2em$\bullet$}%
\vertup\hbox{\kern-0.1em$|$}%
\vertup\hbox{\kern-0.1em$|$}%
\vertup\hbox{\kern -0.2em$\bullet$}%
\kern -2ex\hbox{\kern-0.2em$\scriptstyle #3$}}}%
\kern-0.27em
\raise 1 ex \hbox{$\scriptstyle \, #2$} \kern-0.5 em}
\def\coluubu#1#2#3{\hbox{\raise -6ex \vbox{\normalbaselineskip=3.46 ex 
\baselineskip=3.46 ex 
\nointerlineskip%
\hbox{\kern -0.21em$\buildrel{#1}\over\bullet$\kern -0.2em}
\vertup\hbox{\kern-0.1em$|$}%
\vertup\hbox{\kern-0.1em$|$}%
\vertup\hbox{\kern-0.2em$\uparrow$}%
\vertup\hbox{\kern -0.2em$\bullet$}%
\vertup\hbox{\kern-0.1em$|$}%
\vertup\hbox{\kern-0.1em$|$}%
\vertup\hbox{\kern-0.2em$\uparrow$}%
\vertup\hbox{\kern -0.2em$\bullet$}%
\kern -2ex\hbox{\kern-0.2em$\scriptstyle #3$}}}%
\kern-0.27em
\raise 1 ex \hbox{$\scriptstyle \, #2$} \kern-0.5 em}
\def\coldubu#1#2#3{\hbox{\raise -6ex \vbox{\normalbaselineskip=3.46 ex 
\baselineskip=3.46 ex 
\nointerlineskip%
\hbox{\kern -0.21em$\buildrel{#1}\over\bullet$\kern -0.2em}
\vertup\hbox{\kern-0.2em$\downarrow$}%
\vertup\hbox{\kern-0.1em$|$}%
\vertup\hbox{\kern-0.1em$|$}%
\vertup\hbox{\kern -0.2em$\bullet$}%
\vertup\hbox{\kern-0.1em$|$}%
\vertup\hbox{\kern-0.1em$|$}%
\vertup\hbox{\kern-0.2em$\uparrow$}%
\vertup\hbox{\kern -0.2em$\bullet$}%
\kern -2ex\hbox{\kern-0.2em$\scriptstyle #3$}}}%
\kern-0.27em
\raise 1 ex \hbox{$\scriptstyle \, #2$} \kern-0.5 em}
\def\colddbu#1#2#3{\hbox{\raise -6ex \vbox{\normalbaselineskip=3.46 ex 
\baselineskip=3.46 ex 
\nointerlineskip%
\hbox{\kern -0.21em$\buildrel{#1}\over\bullet$\kern -0.2em}
\vertup\hbox{\kern-0.2em$\downarrow$}%
\vertup\hbox{\kern-0.1em$|$}%
\vertup\hbox{\kern-0.1em$|$}%
\vertup\hbox{\kern -0.2em$\bullet$}%
\vertup\hbox{\kern-0.2em$\downarrow$}%
\vertup\hbox{\kern-0.1em$|$}%
\vertup\hbox{\kern-0.1em$|$}%
\vertup\hbox{\kern -0.2em$\bullet$}%
\kern -2ex\hbox{\kern-0.2em$\scriptstyle #3$}}}%
\kern-0.27em
\raise 1 ex \hbox{$\scriptstyle \, #2$} \kern-0.5 em}
\def\coludbu#1#2#3{\hbox{\raise -6ex \vbox{\normalbaselineskip=3.46 ex 
\baselineskip=3.46 ex 
\nointerlineskip%
\hbox{\kern -0.21em$\buildrel{#1}\over\bullet$\kern -0.2em}
\vertup\hbox{\kern-0.1em$|$}%
\vertup\hbox{\kern-0.1em$|$}%
\vertup\hbox{\kern-0.2em$\uparrow$}%
\vertup\hbox{\kern -0.2em$\bullet$}%
\vertup\hbox{\kern-0.2em$\downarrow$}%
\vertup\hbox{\kern-0.1em$|$}%
\vertup\hbox{\kern-0.1em$|$}%
\vertup\hbox{\kern -0.2em$\bullet$}%
\kern -2ex\hbox{\kern-0.2em$\scriptstyle #3$}}}%
\kern-0.27em
\raise 1 ex \hbox{$\scriptstyle \, #2$} \kern-0.5 em}
\def\upbu#1#2{\hbox{\vtop{\normalbaselineskip=3.46 ex 
\baselineskip=3.46 ex 
\nointerlineskip%
\hbox{\kern -0.21em$\buildrel{#1}\over\bullet$\kern -0.2em}
\vertup\hbox{\kern-0.1em$|$}%
\vertup\hbox{\kern-0.2em$\uparrow$}%
\vertup\hbox{\kern -0.2em$\bullet$}%
\kern -2ex\hbox{\kern-0.2em$\scriptstyle #2$}}}%
\kern-0.27em}
\def\downbu#1#2{\hbox{\vtop{\normalbaselineskip=3.46 ex 
\baselineskip=3.46 ex 
\nointerlineskip%
\hbox{\kern -0.21em$\buildrel{#1}\over\bullet$\kern -0.2em}
\vertup\hbox{\kern-0.2em$\downarrow$}%
\vertup\hbox{\kern-0.1em$|$}%
\vertup\hbox{\kern -0.2em$\bullet$}%
\kern -2ex\hbox{\kern-0.2em$\scriptstyle #2$}}}%
\kern-0.27em}
\def\udbu#1#2{\hbox{\vtop{\normalbaselineskip=3.46 ex 
\baselineskip=3.46 ex 
\nointerlineskip%
\hbox{\kern -0.21em$\buildrel{#1}\over\bullet$\kern -0.2em}
\vertup\hbox{\kern-0.1em$|$}%
\vertup\hbox{\kern-0.1em$|$}%
\vertup\hbox{\kern -0.2em$\bullet$}%
\kern -2ex\hbox{\kern-0.2em$\scriptstyle #2$}}}%
\kern-0.27em}

\def\lsmall{\small}

\title{Trees}
\author{Henri Epstein\\
Institut des Hautes Etudes Scientifiques,
91440 Bures-sur-Yvette, France}
\date{\dateline}

\begin{document}

\maketitle

\begin{abstract}
An algebraic formalism,  developed with V.~Glaser and R.~Stora
for the study of the generalized retarded
functions of quantum field theory, is used to prove a factorization
theorem which provides a complete description of the generalized retarded
functions associated with any tree graph. Integrating over the variables
associated to internal vertices to obtain the perturbative generalized
retarded functions for interacting fields arising from such graphs
is shown to be possible for a large category of space-times.

\noindent
\hbox to 14cm{\hfill\hbox{\it \small
To the memory of Raymond Stora}
}

\end{abstract}

\section{Introduction}
\label{intro}

The general properties  of generalized retarded $n$-point functions
in general field theory were discovered and studied by several
authors in the 1960's \cite{P,R,S1,S2,B3,A1,AB,B2}. In particular
\cite{B2} gives support properties which lead to the full
primitive domain of analyticity in momentum space. In the 1970's
a new presentation, inspired by perturbation theory, was
described in \cite{EGS1,EGS2}. It is based on an algebraic structure
which has some interest in itself, and is well-adapted to 
not necessarily Minkowskian space-times. This paper presents an
application of this algebraic formalism -- for which Raymond Stora
always had a liking -- to a small problem in perturbation theory.
In perturbation theory
the time-ordered or retarded functions of interacting fields
are obtained by integrating graphs in the variables attached to some
internal vertices (see Sect. \ref{perth} for more details). 
In Minkowski space-time 
this integration is always feasible in the absence of
zero masses (see e.g. \cite{EG}), 
but not always when they occur. This is even true in
the case of tree graphs. It was remarked by J.~Bros, in the few-vertex
case, that it is still possible to obtain the retarded function associated
to a tree graph. Here we will see (Sects. \ref{actree} and \ref{perth})
that the generalized retarded functions
for the interacting fields associated to a tree graph can always be
obtained by integrating the variables attached to the internal vertices
over a bounded region, provided one deals with a space-time
in which the double-cones are bounded. This is based on a factorization
theorem (Corollary \ref{grffac}) which follows naturally 
from the algebraic formalism mentioned
above, and which also yields a complete description of the
generalized retarded functions associated to any tree graph
(Sect. \ref{actree}).

Let $\XX$ denote a ``space-time''. This is a smooth manifold
which can be the $d$-dimensional
Minkowski space ($M_d$), or the $d$-dimensional de Sitter space ($dS_d$), or
the universal cover of the $d$-dimensional Anti-de Sitter space 
($\wt{AdS_d}$), or even a more general space-time. 
We suppose that a closed reflexive relation denoted $x \le y$ (or equivalently
$y \ge x$)  is defined in $\XX$. This need not be an order relation,
but it is in the three examples mentioned above. The relation
$x \rle y$ (or equivalently $y \rge x$) is defined as the negation
of $y \le x$. If $A$ and $B$ are subsets of $\XX$, we denote
\beq
A \rle B \Longleftrightarrow B\rge A \Longleftrightarrow
( \forall x \in A\ \ \forall y \in B,
\ \ x \rle y)\ . 
\label{i.2}\endq
Thus (\ref{i.2}) means that there is no $x \in A$ and no $y \in B$
such that $x \ge y$. 
If $x\in \XX$, the future (resp. past) set of $x$ is the (closed) set of all
$y$ such that $y \ge x$ (resp. $y\le x$). If $A\subset \XX$, the future
(resp. past) set of $A$ is the union of the future (resp. past) sets
of all the elements of $A$. The condition (\ref{i.2}) means that
$B$ does not intersect the past set of $A$, or, equivalently,
that $A$ does not intersect the future set of $B$.
If $x, y \in \XX$ the set $\{z\in \XX\ :\ y \le z \le x\}$ is called
the double-cone with vertices $x$ and $y$. If $\le$ is an order relation,
it is empty unless $y\le x$. 
In Minkowski space, and in de Sitter space $dS_d$
viewed as a hyperboloid imbedded in a $(d+1)$-dimensional Minkowski space, 
$x \le y \Leftrightarrow y\in x+\ovl{V_+}$, and 
$(A \rle B) \Longleftrightarrow (A\cap (B + \ovl{V_+}) = \emptyset)$,
where, as usual, $V_+ = \{(x^0,\ \vec{x})\ :\ x^0 > |\vec{x}|\} = -V_-$
and $\ovl{V_+} = \{(x^0,\ \vec{x})\ :\ x^0 \ge |\vec{x}|\}= -\ovl{V_-}$.
In the Minkowski and de Sitter spaces,
the double-cones are compact.

Let $X$ be a finite
set of (distinct) indices with cardinal denoted $|X|$. By $\XX^X$ we
denote $\XX^{|X|}$ or more precisely the set of maps $X \rightarrow \XX$.
$\PP(X)$ denotes the set of subsets of $X$. $\PP_*(X)$ denotes the set
of proper subsets of $X$ i.e. 
$\PP_*(X) = \{J\subset X\ :\ J\not= \emptyset\ \ {\rm and}\ \ J\not= X\}$.
A proper sequence in $\PP(X)$ is a sequence $\{J_1,\ldots,J_\nu\}$
of disjoint non-empty subsets of $X$ with union $X$.
A linear system of generalized time-ordered functions
(GTOF) in variables indexed by $X$ is a set of distributions on $\XX^X$, 
indexed by the proper sequences in $\PP(X)$, usually denoted
\beq
t_{J_1,\ldots,J_\nu} \ \ \ {\rm or}\ \ \ t^c_{J_1,\ldots,J_\nu}\ \ 
\hbox{(connected, or truncated version)}
\label{i.3}\endq
and having the property of being symmetric in the variables with
indices contained in any given $J_k$. In addition we suppose that
for any $k \in X$ these distributions are $\CC^\infty$ in the
variable $x_k$ when smeared with smooth test-functions in the
remaining variables (let us call this the property of partial
regularity, or PR). In the Minkowski case this is a consequence
of the translational invariance which we will always impose, together 
with the usual spectral assumptions, on these
distributions. In the de Sitter or Anti-de Sitter case, PR is a consequence
of the group invariance if it holds. In more general space-times it
follows from the microlocal properties imposed by several authors
as a substitute for translational invariance (see \cite{BF1} and 
references therein).

Last but not least, the set of distributions
$\{t_{J_1,\ldots,J_\nu}\}$ must have the property of {\bf causal factorization}.
This means that if $X=A\cup B$, $A\cap B = \emptyset$, then
\beq
t_{J_1,\ldots,J_\nu} - t_{J_1\cap A,J_1\cap B\ldots,J_\nu\cap A, J_\nu\cap B}\ \ \ 
\hbox{vanishes in the open set}\ \ 
\{x \in \XX^X\ :\ \{x\}_A \rge \{x\}_B\}\ .
\label{i.4}\endq
The last notation means the set of all $x \in \XX^X$ such that
$x_j \rge x_k$ for all $j\in A$ and all $k\in B$.
(If $K$ is a set of indices, $\{x\}_K$ denotes $\{x_j\ :\ j\in K\}$.) 
Of course the term
``causal factorization'' does not mean that the distribution
$t_{J_1,\ldots,J_\nu}$ actually factorizes, but refers to the 
property of causal factorization possessed by time-ordered
products of local quantum fields. If time-ordered products
$T(\{\phi_k(x_k)\}_{k\in K})$, abbreviated as $T(K)$, have been defined,
and $I\cup J = K$, $I\cap J = \emptyset$ then
$T(K)-T(I)T(J)$ vanishes in $\{x\ :\ \{x\}_I \rge \{x\}_J\}$, and the
distributions
\beq
t_{J_1,\ldots,J_\nu} = (\Omega,\ T(J_1)\ldots T(J_\nu)\Omega)
\label{i.4.1}\endq
or their truncated versions 
\beq
t^c_{J_1,\ldots,J_\nu} = (\Omega,\ T(J_1)\ldots T(J_\nu)\Omega)_c
\label{i.4.2}\endq
are examples of GTOF.
See e.g. \cite{EGS2}.
However this paper is concerned with GTOF which exhibit another type of
(actual) factorization, described in the next subsection.

\subsection{Factorization at an index}
\label{fac1}
\begin{figure}[h]
\begin{center}
\setlength{\unitlength}{1 cm}
\includegraphics[scale=0.8]{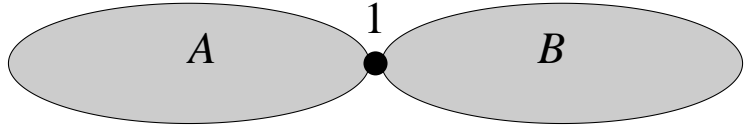}
\caption{Factorization at an internal vertex}
\label{treef}
\end{center}
\end{figure}

We consider a linear system of GTOF in variables 
indexed by a finite set $X = A\cup B\cup \{1\}$,
with $1 \notin A \not= \emptyset$, $1 \notin B \not= \emptyset$,
$A \cap B = \emptyset$, which factorize as symbolized by 
Figure \ref{treef}.

\vskip 0.5 truecm
This means that the linear system of  GTOF $t^c$
satisfies
\begin{equation}
t^c_{J_1,\ldots,J_\nu}(\{x\}_X) = 
t^{1,c}_{J_1\cap(A\cup 1),\ldots,J_\nu\cap(A\cup 1)}(\{x\}_{A\cup 1})\,
t^{2,c}_{J_1\cap(B\cup 1),\ldots,J_\nu\cap(B\cup 1)}(\{x\}_{B\cup 1})
\label{1.1}\end{equation}
where the $\{t^{1,c}\}$ (resp. $\{t^{2,c}\}$) are a linear system of 
GTOF in variables indexed by  $A\cup\{1\}$
(resp. $B\cup\{1\}$). In the case when the $t^c$ are 
connected (or truncated) vacuum expectation
values of products of time-ordered products of local fields, this
takes the form
\begin{eqnarray}
\lefteqn{
(\Omega,\ T(J_1)\ldots T(J_\nu)\Omega)_c}\nonumber\\
&& =
(\Omega,\ T(J_1\cap(A\cup 1))\ldots T(J_\nu\cap(A\cup 1))\Omega)_{1,c}
\nonumber\\
&& \times
(\Omega,\ T(J_1\cap(B\cup 1))\ldots T(J_\nu\cap(B\cup 1))\Omega)_{2,c}
\label{1.2}\end{eqnarray}

This makes sense because of the PR property: smearing with test-functions in
the variables indexed by $A$ and $B$ yields a product of two
$\CC^\infty$ functions of $x_1$. In addition the product (\ref{1.1}) itself
has the property PR. This is obvious if the distinguished variable is $x_1$.
If the distinguished variable is $x_j$, with $j\in A$, we first smear in
the variables indexed by $B$. The second factor then becomes a $\CC^\infty$
function of $x_1$ and smearing the variables indexed by 
$\{1\}\cup A\setminus \{j\}$ finally yields a $\CC^\infty$ function of $x_j$.
In the Minkowskian case, 
we can take as variables the two independent
groups $\{x_j - x_1\ :\ j \in A\}$ and $\{x_j - x_1\ :\ j \in B\}$, 
so that (\ref{1.1}, \ref{1.2}) are really tensor products of distributions.

The main example of this situation is the system of GTOF associated
to a perturbative {\it tree graph} with vertices labelled by $X$, in which
1 is an internal vertex, and $A\cup \{1\}$ and $B\cup \{1\}$
are the sets of vertices of two subtrees. A more general example is
given by any perturbative graph which is splittable at an internal point
(see again Fig. \ref{treef}).
In this case, 
we are really dealing with Wick monomials of free 
or generalized free fields \cite{BLOT,J,BF1,SW,S4}, 
each field indexed by $j \in A$ being of the form
$:\phi^{m_j}(x_j): \otimes 1$, each field indexed by $k \in B$
being of the form $1 \otimes :\psi^{m_k}(x_k):$, and the field
labelled by 1 being $:\phi^{a}(x_1):\otimes :\psi^b(x_1):$,
operating in a tensor product of two Fock spaces.

\vskip 1 cm

In the Minkowskian case, given a distribution
\begin{equation}
F(\{x'_j\}_{j \in A},\ \{x''_k\}_{k \in B},\ x_1) = 
f(\{x'_j-x_1\}_{j \in A},\ \{x''_k-x_1\}_{k \in B})
\label{1.3}\end{equation}
\begin{equation}
= F_1(\{x'_j\}_{j \in A},\ x_1)\,
F_2(\{x''_k\}_{k \in B},\ x_1)
\label{1.4}\end{equation}
\begin{equation}
=f_1(\{x'_j-x_1\}_{j \in A})\,
f_2(\{x''_k-x_1\}_{k \in B}),
\label{1.5}\end{equation}
we have, for the Fourier transforms:
\begin{equation}
\wt F(\{p'\}_A,\ \{p''\}_B,\ p_1) 
=\delta(p'_A +p''_B +p_1)\, \wt f(\{p'\}_A,\ \{p''\}_B),
\label{1.6}\end{equation}
\begin{equation}
\wt f(\{p'\}_A,\ \{p''\}_B) =
\wt f_1(\{p'\}_A)\, \wt f_2(\{p''\}_B),
\label{1.7}\end{equation}
\begin{equation}
\wt F_1(\{p'\}_A,\ p_1) = 
\delta(p'_A + p_1)\, \wt f_1(\{p'\}_A),
\label{1.8}\end{equation}
\begin{equation}
\wt F_2(\{p''\}_B,\ p_1) = 
\delta(p''_B + p_1)\, \wt f_2(\{p''\}_B),
\label{1.9}\end{equation}
where we denote
\begin{equation}
\{p'\}_A = \{p'_j\}_{j\in A},\ \ \ 
\{p''\}_B = \{p''_k\}_{k\in B},\ \ \ 
p'_J = \sum_{j \in J} p'_j\ .
\label{1.10}\end{equation}

\section{Generalized retarded operators and functions}
\label{grofs}

This section is a summary of a part of \cite{EGS2} which will 
be applied in the subsequent sections to factorizing systems.

\subsection{The algebra of sequences $\AA(X)$}

In this subsection $X$ denotes a finite set such that $|X| \ge 2$.
We denote $\PP(X)$ the set of subsets of $X$.
We denote $\PP_*(X)$ the set of {\bf proper subsets} of
$X$, i.e.
\begin{equation}
\PP_*(X) = \{J \subset X\ :\ J \not= \emptyset,\ \ 
J \not= X\}
\label{2.1}\end{equation}

\begin{definition}
\label{algebra}
A {\bf proper sequence in $\PP(X)$} is a sequence
$\{J_1,\ldots, J_\nu\}$, 
where $\nu = 1,\ 2,\ldots,\ |X|$, the $J_k$ are {\bf non-empty, disjoint}
subsets of $X$ with union $X$. A multiplication law is defined for proper
sequences by
\begin{align}
\{A_1,\ \ldots,\ A_n\}&\{B_1,\ \ldots,\ B_m\}\ = \cr
&\{A_1\cap B_1,\ \ldots,\ A_n\cap B_1,\ \ldots,
\ A_1\cap B_m,\ \ldots,\ A_n\cap B_m\}
\ {\rm mod}\ \emptyset\ .
\label{2.1.1}\end{align}
Here ``mod $\emptyset$'' means: ``omit every occurrence of the empty set''.
The algebra $\AA(X)$ (called the algebra of sequences in $\PP(X)$) is the
vector space of all formal complex linear combinations of proper sequences 
in $\PP(X)$, equipped with the multiplication generated by (\ref{2.1.1}).
\end{definition}
The multiplication is associative and $\AA(X)$ has all the usual 
properties of an algebra. It has a unit, namely $\{X\}$. 
Any proper sequence $c$ in $\PP(X)$ is an idempotent, i.e. $cc=c$.
When no ambiguity arises,
we denote $\wh I = \{I,\ X\setminus I\}$ for any proper subset $I$ of $X$.
Since $\wh I$ is a proper sequence, $\wh I \wh I = \wh I$.

The following lemma is easy to prove (see \cite{EGS2} p.19)

\begin{lemma}
\label{annh}
Let $I \in \PP_*(X)$, $a,\ b \in \AA(X)$, 
$\wh I = \{I,\ X\setminus I\} \in \AA(X)$. If $a$ is such that
$(1-\wh I)\,a =0$, then also $(1-\wh I)\,ba =0$.
\end{lemma}

\begin{definition}
\label{geom}
A {\bf geometrical cell} associated to $X$ is one of the 
connected components of
\begin{equation}
\{s = \{s\}_X \in \bR^{|X|}\ :\ 
s_X = 0,\ \ \ s_J \not= 0\ \ \forall J \in \PP_*(X)\}
\label{2.2}\end{equation}
Here $\{s\}_X$ denotes the set of variables $\{s_j\ :\ j \in X\}$,
and, for every subset $J$ of $X$, $s_J \bydef \sum_{j\in J}s_j$.
\end{definition}

A picture of the geometrical cells for $X=\{1,\ 2,\ 3,\ 4\}$
is given in Figure \ref{stplanet} in 
Appendix 
\ref{app1}.

\begin{definition}
\label{paracell}
A {\bf paracell} associated to $X$ is a non-empty subset $\SS$ of $\PP_*(X)$
such that if $I \in \SS$ and $J \in \SS$, then $I\cap J \in \SS$ or
$I \cup J \in \SS$. In this case 
$\SS' = \{J \subset X\ :\ X \setminus J \in \SS\}$
is also a paracell, and $\SS \cap \SS' = \emptyset$. 
The two paracells $\SS$ and $\SS'$
are said to be opposite.
\end{definition}
There are no paracells associated to $X$ unless
$|X| \ge 2$.

It follows from this definition that if $I \in \SS$ and $J \in \SS$,
then $I \cap J = \emptyset \Rightarrow I \cup J \in \SS$, and
$I \cup J = X \Rightarrow I \cap J\in \SS$.

\begin{definition}
\label{precell}
A {\bf precell} associated to $X$ is a paracell $\SS$ associated to $X$
such that, for every $J \in \PP_*(X)$, $J \in \SS$ or $X \setminus J \in \SS$.
In other words $\SS \cup \SS' = \PP_*(X)$,
where $\SS'$ denotes the paracell opposite to $\SS$. 
In this case $\SS'$ is also a precell.
\end{definition}

The following are trivial consequences of the definition,
assembled in a lemma for future reference.

\begin{lemma}
\label{precell2}
Let $\SS$ be a precell.

(i) if $J \in \SS$ then $X \setminus J \notin \SS$~;

(ii) if $J \in \PP_*(X)$ and $J \notin \SS$ then $X \setminus J \in \SS$~;

(iii) if $J\in \SS$, $K\in \SS$, and $J \cap K = \emptyset$, 
then $J\cup K \in \SS$;

(iv) if $J\in \SS$, $K\in \SS$, and $J \cup K = X$, then $J\cap K \in \SS$.

(v) if $K \cup L = J \in \SS$ and $K\cap L = \emptyset$, then $K \in \SS$
or $L \in \SS$.

Conversely, if $\SS$ is a subset of $\PP_*(X)$ having the properties
(i)-(iv), and if $\SS'$ denotes $\{J \subset X\ :\ X\setminus J \in \SS\}$,
then $\SS$ and $\SS'$ are opposite precells relative to $X$.
\end{lemma}

To prove (v) we note that the assertion is obvious if either
$K$ or $L$ is empty. Otherwise they must both belong to $\PP_*$,
and cannot both belong to $\SS'$ since that would imply
$J \in \SS'$, hence at least one of them belongs to $\SS$.
Note that this assertion does not hold for general paracells.

\begin{definition}
\label{cell}
A cell associated to $X$ is a precell $\SS$
such that there exist $|X|$ real numbers $\{s_j\ :\ j \in X\}$
satisfying
\begin{equation}
\sum_{j \in X} s_j = 0,\ \ \ 
s_J\ \bydef\ \sum_{j \in J} s_j > 0\ \ \ \forall J \in \SS.
\label{2.2.1}\end{equation}
In other words a cell associated to $X$ is a precell $\SS$
such that there is a geometrical cell
$\CC_\SS$ such that, for every $J \in \SS$, and every $s \in \CC_\SS$,
$s_J > 0$.
\end{definition}

$\SS$ and $\CC_\SS$ are then uniquely determined by each other
and $\SS'$ is the cell associated to the geometrical cell
opposite to $\CC_\SS$, i.e. $\CC_{\SS'} = -\CC_\SS$. It is clear that
every geometrical cell determines a cell in this way.

If $Y$ and $X$ are disjoint non-empty finite sets, and $\SS$ and $\SS'$
are opposite paracells associated to $X$, we denote
\begin{eqnarray}
Y \downarrow \SS &=& \{J \in \PP_*(Y\cup X)\ :\ 
J\cap X = X\ {\rm or}\ J\cap X \in \SS\}\label{2.2.2}\\
Y \uparrow \SS' &=& \{J \in \PP_*(Y\cup X)\ :\ 
J\cap X = \emptyset\ {\rm or}\ J\cap X \in \SS'\}
\label{2.2.3}\end{eqnarray}
It is easily verified that $Y \downarrow \SS$ and
$Y \uparrow \SS'$ are opposite paracells associated to
$Y\cup X$.
If $\SS$ and $\SS'$ are opposite precells
(resp. cells) associated to $X$, then $Y \downarrow \SS$ and 
$Y \uparrow \SS'$ are opposite precells 
(resp. cells) associated to $X\cup Y$.
In case $Y$ has only one element $j$ we abbreviate 
$\{j\} \updownarrow \SS$ to $j \updownarrow \SS$.

If $X$ has only one element (denoted $j$), there are no
paracells associated to $X$, but we define
\begin{eqnarray}
Y \downarrow j &=& \{J \in \PP_*(Y\cup X)\ :\ j \in J\},\label{2.2.2.1}\\
Y \uparrow j &=& \{J \in \PP_*(Y\cup X)\ :\ j\notin J\}.
\label{2.2.3.1}\end{eqnarray}
$Y\downarrow j$ and $Y \uparrow j$ are opposite cells, corresponding
respectively to the geometrical cell
\begin{equation}
\{s = \{s\}_{Y\cup X}\in \bR^{|Y|+1}\ :\ 
s_j + \sum_{k \in Y} s_k = 0,\ \ 
s_k < 0\ \forall k \in Y\}
\label{2.2.4.1}\end{equation}
and its opposite.

If $Y = (j_1\,,\ j_2\,,\ \ldots\ ,\ j_p)$,
\beq
Y\uparrow \SS = j_1\uparrow j_2\uparrow \ldots j_p\uparrow \SS,\ \ \ \ 
Y\downarrow \SS = j_1\downarrow j_2\downarrow \ldots j_p\downarrow \SS.
\label{2.2.4.2}\endq
This is independent of the order of the $j_k$'s.

If $|X| > 1$ and $j \in X$, then $\{j\} \in Y\uparrow \SS$ 
(resp. $\{j\} \in Y\downarrow \SS$) if and only if $\{j\} \in \SS$.
If $X= \{j\}$ then $\{j\} \in Y\downarrow j$ and 
$\{j\} \not\in Y\uparrow j$.

If $X$ has two elements, say $X= (1,\ 2)$ then the only paracells
associated to $X$ are $\{\{1\}\} = 1\uparrow 2 = 2\downarrow 1$ 
and its opposite $\{\{2\}\} = 1\downarrow 2 = 2\uparrow 1$. 
They are cells corresponding respectively to
the geometrical cells 
\beq
\{s_1,\ s_2\ :\ s_1+s_2 =0,\ \ s_1 >0\}\ \ \hbox{and}\ \ 
\{s_1,\ s_2\ :\ s_1+s_2 =0,\ \ s_1 <0\}\ .
\label{2.2.4.3}\endq
Cells of the form $j_1\updownarrow\ldots j_{k-1}\updownarrow j_k$
are called Steinmann monomials (there are cells which are not
of this form).
In fact the operations $Y\uparrow$ and $Y\downarrow$ can be defined
as linear maps of the whole algebra $\AA(X)$ into $\AA(Y\cup X)$. See
\cite{EGS2} for details.

\begin{definition}
\label{chains}
Let $\SS$ be a paracell in $X$, and $\SS'$ the opposite paracell.
A $\nu$-chain associated to $\SS'$ is a sequence
$\{J_1,\ \ldots,\ J_\nu\}$ of $\nu$ disjoint, non-empty subsets
of $X$ such that $J_1\cup \ldots\cup J_\nu =X$ and that, for
every $r < \nu$,
\begin{equation}
I_r \ \bydef\ J_1\cup \ldots\cup J_r \in \SS'\ .
\label{2.3}\end{equation}
This requires $1 \le \nu \le |X|$, and, of course, if $\nu = 1$
the condition (\ref{2.3}) is empty.

We denote $\MM_\nu(\SS')$ the set of all $\nu$-chains associated to $\SS'$.
In particular $\MM_1(\SS') = \{\{X\}\}$.
\end{definition}
Thus, in particular, $\MM_\nu(\SS') \subset \AA(X)$. 
For any paracell $\SS$, with opposite paracell $\SS'$, we denote
\begin{equation}
\UU_\SS = \sum_{\nu=1}^{|X|}\ 
\sum_{\{J_1,\ldots,J_\nu\} \in \MM_\nu(\SS')}
(-1)^{\nu-1}\,\{J_1,\ldots,J_\nu\}
\label{2.3.1}\end{equation}

The following lemmas are proved in \cite{EGS2}.
\begin{lemma}
\label{parorder}
Let $\SS'$ be a paracell in $X$, and $\{I_1,\ldots,\ I_N\}$ be an arbitrary
ordering of all the elements of $\SS'$. Then, for any permutation
$\pi$ of $(1,\ldots,\ N)$,
\begin{equation}
(1-\wh I_1)\ldots (1- \wh I_N) =
(1-\wh I_{\pi(1)})\ldots (1- \wh I_{\pi(N)}).
\label{2.3.2}\end{equation}
In other words the lhs of (\ref{2.3.2}) does not depend on the chosen
ordering.
\end{lemma}

\begin{lemma}
\label{isplit}
Let $\SS$ be a paracell with opposite paracell $\SS'$. Then
\begin{equation}
\forall I \in \SS'\ \ \ \wh I \UU_\SS = 0.
\label{2.3.3}\end{equation}
\end{lemma}
It follows that
\beq
\forall I \in \SS'\ \ \ (1-\wh I) \UU_\SS = \UU_\SS\ .
\label{2.3.4}\endq

\begin{lemma}
\label{uform}
With the notations of Lemma \ref{parorder},

\begin{equation}
\UU_\SS = (1-\wh I_1)\ldots (1- \wh I_N)
\label{2.3.5}\end{equation}
\end{lemma}

\subsection{Generalized retarded operators and functions}
\label{grf}
Given a set of time-ordered products for fields indexed by $X$,
(resp. a linear system of GTOF indexed by
$X$), we can define a linear map of the algebra $\AA(X)$ into
the operator-valued distributions (resp. the distributions) by
defining, for each proper sequence $a = \{J_1,\ldots,J_\nu\}$
\begin{equation}
\mbox{\bf T}a = T(J_1)\ldots T(J_\nu),
\label{2.3.6}\end{equation}
\begin{equation}
{\rm resp.}\ \ \ \mbox{\bf t}a = t^c_{J_1,\ldots,J_\nu}\ .
\label{2.3.7}\end{equation}
This extends by linearity to the whole $\AA(X)$ since the
proper sequences are (by definition) a basis of this vector
space. If $b$ is an element of $\AA(X)$, we will also denote
$t_b^c = \mbox{\bf t}b$.

In particular we may associate to every paracell $\SS$,
with opposite paracell $\SS'$
\begin{equation}
R_{\SS} = \mbox{\bf T}{\UU_\SS} =
\sum_{\nu=1}^{|X|}(-1)^{\nu-1}
\sum_{\{J_1,\ldots,J_\nu\} \in \MM_\nu(\SS')}
T(J_1)\ldots T(J_\nu)
\label{2.4}\end{equation}
resp.
\begin{equation}
r_{\SS} =\mbox{\bf t}\,{\UU_\SS} =
\sum_{\nu=1}^{|X|}(-1)^{\nu-1}
\sum_{\{J_1,\ldots,J_\nu\} \in \MM_\nu(\SS')}
t^c_{J_1,\ldots,J_\nu}\ 
\label{2.5}\end{equation}
In the special case when $\SS$ and $\SS'$ are cells,
the $R_\SS$ (resp. $r_\SS$) are the generalized retarded operators
(resp. generalized retarded functions, abbreviated to GRF).
In this special case, in
the last equation, the $t^c_{J_1,\ldots,J_\nu}$ can be {\it all}
replaced by their non-truncated versions without affecting the result
(in other words the GRF are naturally truncated. For this 
well-known fact see \cite{A1,AB,B2,B3,EGS2,R,S1,S2}).

The property of causal factorization (\ref{i.4}) can be reexpressed
as follows: if $I'$ is a proper subset of $X$ and $a$ is a proper sequence
in $\PP(X)$, then 
\beq
\mbox{\bf T}\,(1-\wh I')a \ \hbox{(resp. {\bf t}$\,(1-\wh I')a$)}\ \ 
\hbox{vanishes in the open set}\ \ 
\{x \in \XX^X\ :\ \{x\}_{I'} \rge \{x\}_{X\setminus I'}\}\ .
\label{s.0.0}\endq
This, of course, extends to any $a\in \AA(X)$.
For every precell $\SS$ opposite to $\SS'$, and every $I' \in \SS'$,
the identity 
$(1-\wh I')\UU_\SS= \UU_\SS$ and eqs. (\ref{2.4}), (\ref{2.5}),
and (\ref{s.0.0}) imply that
\begin{equation}
R_\SS \ {\rm and}\ r_\SS\ {\rm vanish\ in}\ 
\bigcup_{I \in \SS} \{\,\{x\}_X\ :\ \{x\}_I \rle \{x\}_{X \setminus I}\},
\label{s.0.1}\end{equation}
or, equivalently,
\begin{equation}
{\rm support}\ R_\SS \ {\rm and\ support}\ r_\SS \subset
\bigcap_{I \in \SS} \{\,\{x\}_X\ :\ 
\exists j \in I,\ \exists k \in X \setminus I\ \ {\rm s.t.}\ 
x_k \le x_j \}.
\label{s.0.2}\end{equation}
As the simplest example, if $X = (1,\ 2)$ the only paracells
are, as we have seen, $\SS = \{\{1\}\} = 1\uparrow 2 = 2\downarrow 1$ and
$\SS' = \{\{2\}\} = 1\downarrow 2 = 2\uparrow 1$, and
\beq
{\rm support}\ R_{1\uparrow 2} \subset
\{(x_1,\ x_2)\ :\ x_1 \ge x_2\},\ \ \ 
{\rm support}\ R_{1\downarrow 2} \subset
\{(x_1,\ x_2)\ :\ x_1 \le x_2\}\ .
\label{s.0.3}\endq
Of course the same holds for the $r_\SS$.

\section{Factorization}
\label{facthm}
\subsection{Some algebra}
\label{somalg}
In applying Sect.~\ref{grofs} to the case of a 
linear system of time-ordered functions
which factorize as in Subsect.~\ref{fac1}, there is a first 
non-trivial fact to prove, namely that, given a cell $\SS$
relative to $X= A\cup B\cup \{1\}$, there are cells
$\SA$ relative to $A\cup\{1\}$ (opposite to
$\SPA$) and $\SB$ relative to $B\cup\{1\}$ 
(opposite to $\SPB$), such that
\begin{equation}
r_{\SS} = r_{\SA}\,r_{\SB},
\label{3.1}\end{equation}
and similarly with operators, if we make the suitable commutation assumptions.

\vskip 1cm
We suppose, as above, that $X = A \cup B \cup \{1\}$,
with $A$ and $B$ disjoint and non-empty and $1 \notin A\cup B$.
The tensor product $\AA(A\cup \{1\}) \otimes \AA(B\cup \{1\})$
is defined in the standard way, i.e. if $c,\ c'$ are proper sequences
in $\PP(A\cup \{1\})$, $f,\ f'$ are proper sequences in $\PP(B\cup \{1\})$,
then $(c \otimes f)(c' \otimes f') = (cc')\otimes (ff')$, and this extends
by linearity.
We define a linear map $\mbox{\bf Fac}$ of $\AA(X)$ into
$\AA(A\cup \{1\}) \otimes \AA(B\cup \{1\})$ by defining
 $\mbox{\bf Fac}(c)$ for an arbitrary proper sequence
$c = \{J_1,\ldots,\ J_\nu\}$ in $\PP(X)$ as follows:
\begin{eqnarray}
\mbox{\bf Fac}(c) &=& c_A \otimes c_B\ ,\nonumber\\
c_A &=& \{J_1\cap(A\cup \{1\}),\ \ldots,\ J_\nu\cap(A\cup \{1\})\}
\ {\rm mod}\ \emptyset,
\nonumber\\
c_B &=& \{J_1\cap(B\cup \{1\}),\ \ldots,\ J_\nu\cap(B\cup \{1\})\}
\ {\rm mod}\ \emptyset\ .
\label{3.2}\end{eqnarray}
Again this extends by linearity to all of $\AA(X)$. From this formula
and the definition of the multiplication in $\AA(X)$ it immediately
follows that if $c$ and $f$ are two proper sequences in $\PP(X)$,
with $\mbox{\bf Fac}(c) = c_A \otimes c_B$, 
$\mbox{\bf Fac}(f) = f_A \otimes f_B$, then 
$\mbox{\bf Fac}(cf) = c_Af_A \otimes c_Bf_B$, and hence
\begin{equation}
\mbox{\bf Fac}(cf) = \mbox{\bf Fac}(c)\, \mbox{\bf Fac}(f)
\label{3.2.1}\end{equation}
holds for any $c,\ f \in \AA(X)$.

\begin{lemma}
Let $\SS$ and $\SS'$ be two opposite {\bf precells} relative to $X$. 
We denote
\begin{equation}
\SA = \{J \in \PP_*(A\cup \{1\})\ :\ 
J \subset A,\ J\in \SS\}\cup
\{J \in \PP_*(A\cup \{1\})\ :\ 
1 \in J,\ A\setminus J \in \SS'\}\ ,
\label{3.3}\end{equation}
\begin{equation}
\SPA = \{J \in \PP_*(A\cup \{1\})\ :\ 
J \subset A,\ J\in \SS'\}\cup
\{J \in \PP_*(A\cup \{1\})\ :\ 
1 \in J,\ A\setminus J \in \SS\}\ ,
\label{3.4}\end{equation}
\begin{equation}
\SB = \{J \in \PP_*(B\cup \{1\})\ :\ 
J \subset B,\ J\in \SS\}\cup
\{J \in \PP_*(B\cup \{1\})\ :\ 
1 \in J,\ B\setminus J \in \SS'\}\ ,
\label{3.5}\end{equation}
\begin{equation}
\SPB = \{J \in \PP_*(B\cup \{1\})\ :\ 
J \subset B,\ J\in \SS'\}\cup
\{J \in \PP_*(B\cup \{1\})\ :\ 
1 \in J,\ B\setminus J \in \SS\}\ .
\label{3.6}\end{equation}
Then 

(i) $\SA$ and 
$\SPA$ are opposite precells relative to
$A\cup \{1\}$, and $\SB$ and
$\SPB$ are opposite precells relative to
$B\cup \{1\}$;

(ii) If $\SS$ is a cell then $\SA$, $\SPA$, $\SB$ and $\SPB$ are cells.
\end{lemma}

{\bf Proof}. (i) It suffices to prove that under the hypotheses
of the lemma $\SA$ and $\SPA$ are opposite precells relative to
$A\cup \{1\}$. Suppose first that $J$ and $J'$ are complementary
proper subsets of $A\cup \{1\}$. If $J \in \SA$ it is immediate
that $J' \in \SPA$, and that $J' \notin \SA$. If $J\notin \SA$
and $J \subset A$ then $J\notin \SS$ so $J\in \SS'$ hence $J\in\SPA$.
If $J\notin \SA$ and $1 \in J$, then $J'\subset A$ and $J' \notin \SS'$, 
hence $J'\in\SS$, hence again $J\in\SPA$. Thus $\SA$ possesses
the properties (i) and (ii) of Lemma \ref{precell2}. Assume now
that $J\in \SA$, $K\in\SA$ and $J\cap K=\emptyset$. 
Let $J' = (A\cup\{1\})\setminus J$ and $K' = (A\cup\{1\})\setminus K$
If $J\subset A$ and $K\subset A$ then $J\in \SS$, $K\in \SS$
and $J\cup K \in \SS$ and $J\cup K \subset A$, so that $J\cup K \in \SA$.
If $J\subset A$ and $1 \in K$ then $J\in \SS$ and $J\subset K'\in \SS'$.
Therefore $K'\setminus J = K'\cap J' \in \SS'$ hence $K'\cap J' \in \SPA$
which, as we have seen before, implies $J\cup K\in \SA$. Thus $\SA$ possesses
the property (iii) of Lemma \ref{precell2}. So does $\SPA$ since its
definition is symmetrical to that of $\SA$, and this implies
that both of them possess the property (iv). By Lemma \ref{precell2},
this finishes the proof of (i).

\noindent (ii) Let $s$ be a point of the geometrical cell associated
to $\SS$. Let $s'_j = s_j$ for every $j\in A$, $s'_1 = s_1+s_B$.
then $s'_{A\cup\{1\}}=0$, $s'_J >0$ for every $J\subset A$ such that
$J \in \SS$, or every $J \subset A\cup\{1\}$ such that $1\in J$ and
$A\setminus J \in \SS'$, i.e. for every $J \in \SA$. Therefore
$s'_K <0$ for every $K \in \SPA$. Since this accounts for all
proper subsets of $A\cup\{1\}$, $\SA$ and $\SPA$ are opposite cells, 
and similarly for $\SB$ and $\SPB$.\ \endprf

We will prove:

\begin{theorem}
\label{algfac}
Under the above assumptions (including in particular $\SS$ and $\SS'$
being opposite precells)
\begin{equation}
\mbox{\rm \bf Fac} (\UU_\SS) =
\UU_{\SA} \otimes \UU_{\SB}\ .
\label{3.7}\end{equation}
\end{theorem}

{\bf Proof}.
Let $K \in \SPA$, and let 
$K' = (A\cup \{1\}) \setminus K$. We claim that
$(\{K,\ K'\}\otimes 1)\, \mbox{\bf Fac}(\UU_\SS) = 0$. There are
two cases to consider.

{\it Case 1}:
$K \subset A$ and $K \in \SS'$.
Let $I = K \in \SS'$, $I' = X\setminus I$. Then 
$\mbox{\rm \bf Fac} (\{I,\ I'\}) = \{K,\ K'\}\otimes 1$.
By Lemma \ref{isplit},
$\{I,\ I'\}\UU_\SS =0$, and applying $\mbox{\bf Fac}$
gives $(\{K,\ K'\}\otimes 1)\, \mbox{\bf Fac}(\UU_\SS) = 0$.

{\it Case 2}:
$1 \in K$ and $K' \in \SS$. Let $I = X\setminus K' \in \SS'$.
Again $(\{K,\ K'\}\otimes 1)\, \mbox{\bf Fac}(\UU_\SS) =
\mbox{\bf Fac}(\{I,\ X\setminus I\}\UU_\SS) =0$.

Similarly, for $L \in \SPB$, $L' = (B\cup \{1\}) \setminus L$, we have
$(1\otimes\{L,\ L'\})\, \mbox{\bf Fac}(\UU_\SS) = 0$.
As a consequence
\begin{equation}
\mbox{\bf Fac}(\UU_\SS) =
\prod_{K \in \SPA}(1 - \{K,\ K'\}) \otimes
\prod_{L \in \SPB}(1 - \{L,\ L'\})\,
\mbox{\bf Fac}(\UU_\SS)\ .
\label{3.8}\end{equation}
Here $K'$ stands for $A\cup\{1\}\setminus K$, and
$L'$ stands for $B\cup\{1\}\setminus L$. Note that the order
of factors in the two products is irrelevant by Lemma \ref{parorder}.
Let $(-1)^{\nu-1} c = (-1)^{\nu-1}\{J_1,\ldots,\ J_\nu\}$ be
one of the terms in the expansion (\ref{2.3.1}) of $\UU_\SS$ with $\nu > 1$.
Then $J_1 \in \SS'$. This implies that
$K = J_1 \cap (A\cup \{1\}) \in \SPA$
or $L = J_1 \cap (B\cup \{1\}) \in \SPB$.
Indeed suppose first that $1 \notin J_1$. Then 
$J_1\cap (A\cup \{1\}) = J_1 \cap A$ and $J_1\cap (B\cup \{1\}) = J_1 \cap B$
are disjoint subsets with union $J_1 \in \SS'$.
At least one of them must belong to $\SS'$ (Lemma~\ref{precell2} (v)).
Suppose now that $1 \in J_1$.
Then $K' = (A\cup \{1\})\setminus K = A\setminus K$ and
$L' = (B\cup \{1\})\setminus L = B\setminus L$ are disjoint
sets with union $X\setminus J_1 \in \SS$, and at least one of them
belongs to $\SS$. Therefore $(1-\{K,\ K'\})$ occurs
in $\prod_{K \in \SPA}(1 - \{K,\ K'\})$
or $(1-\{L,\ L'\})$ occurs in
$\prod_{L \in \SPB}(1 - \{L,\ L'\})$.
Suppose e.g. that $K = J_1 \cap (A\cup \{1\}) \in \SPA$ and hence
$(1-\{K,\ K'\})$ occurs
in $\prod_{K \in \SPA}(1 - \{K,\ K'\})$.
We have
\beq
c_A = \{K,\ J_2\cap(A\cup \{1\}),\ldots,\ J_\nu\cap(A\cup \{1\})\}
\mod\ \emptyset.
\label{3.8.1}\endq
All the sets $J_k\cap(A\cup \{1\})$ with $k>1$ are contained in $K'$
hence $(1 - \{K,\ K'\})c_A = 0$.
Similarly if $L\in \SPB$, $(1 - \{L,\ L'\})c_B = 0$.
Applying Lemma~\ref{annh} (or Lemma~\ref{parorder})
we see that the contribution of $c$ to the rhs of (\ref{3.8})
vanishes. There remains only the contribution of $\{X\}$, and this
proves the theorem.\ \endprf

\begin{remark}
\rm
The above theorem does not extend to the case when $\SS$ and $\SS'$
are arbitrary opposite paracells. For example if $1 \in I \in \PP_*(X)$
and $I' = X\setminus I$, we can define a paracell $\SS = \{I\}$
with $\SS' =  \{I'\}$ as the opposite paracell. Then
$\UU_\SS = \{X\} - \{I',\ I\}$ and
\begin{equation}
\mbox{\bf Fac}(\UU_\SS) = \{A\cup \{1\}\} \otimes \{B\cup \{1\}\}
-\{I'\cap A,\ I\cap (A \cup \{1\})\} \otimes 
\{I'\cap B,\ I\cap (B \cup \{1\})\}.
\label{3.9}\end{equation}
Moreover if we suppose $I \not\subset A$,  $I' \not\subset A$,
$I \not\subset B$,  $I' \not\subset B$, then $\SA$,
$\SPA$, $\SB$, and $\SPB$
as defined by (\ref{3.3}-\ref{3.6}) are empty.

\end{remark}

A system of GTOF which factorizes as in (\ref{1.1}) satisfies
\begin{equation}
\mbox{\bf t}c = (\mbox{\bf t}^1 c_A)(\mbox{\bf t}^2 c_B)
\label{3.10}\end{equation}
for every proper sequence $c \in \AA(X)$, with $c_A$ and $c_B$
given by (\ref{3.2}). This can be reexpressed symbolically as
\begin{equation}
\mbox{\bf t}c = \mbox{\bf t}^1 \otimes \mbox{\bf t}^2\,\mbox{\bf Fac}\,c
\label{3.11}\end{equation}
for every $c \in \AA(X)$. As a consequence,

\begin{corollary}
\label{grffac}
Given a system of GTOF which factorize as in (\ref{1.1}),
the associated GRF also factorize:
\begin{equation}
r_{\SS} = r_{\SA}\,r_{\SB},
\label{3.12}\end{equation}
for every cell $\SS$ relative to $X$, with opposite cell $\SS'$,
and with the notations (\ref{3.3}-\ref{3.6}).
\end{corollary}

For future use we need an explicit description of $\SA$, $\SPA$,
$\SB$, $\SPB$ in the case when
$\SS = 1 \downarrow \TT$, 
$\SS' = 1 \uparrow \TT'$, where $\TT$ and $\TT'$ are two opposite cells 
relative to $A \cup B$. This means
\begin{equation}
\SS = 1 \downarrow \TT = \{J \in \PP_*(X)\ :\ 
J = A\cup B\ {\rm or}\ J\cap(A\cup B) \in \TT\}
\label{h.10}\end{equation}
\begin{equation}
\SS' = 1 \uparrow \TT' = \{J \in \PP_*(X)\ :\ 
J = \{1\}\ {\rm or}\ J\cap(A\cup B) \in \TT'\}
\label{h.11}\end{equation}
According to our definitions,
\begin{align}
\DTA &= 
\{J\subset A\cup\{1\}\ :\ 
J \subset A\ {\rm and}\ J\in \TT\}\cup\cr
&\{J\subset A\cup\{1\}\ :\ 1 \in J\ {\rm and}\ A\setminus J \in \TT'\},
\label{h.12}\end{align}
\begin{align}
\UTPA &= 
\{J\subset A\cup\{1\}\ :\ 
J \subset A\ {\rm and}\ J\in \TT'\}\cup\cr
&\{J\subset A\cup\{1\}\ :\ 1 \in J\ {\rm and}\ A\setminus J \in \TT\},
\label{h.13}\end{align}
and the same with $B$ instead of $A$.

\subsection{Analyticity in momentum space in the Minkowskian
case}
In this subsection\footnote{The contents of this subsection will not be used
in the remainder of this paper.
Readers unfamiliar with the subject of momentum-space
analyticity, or whose memories of it have faded with time, might
omit this subsection (with the possible exception of its
last sentence).}
we assume the $d$-dimensional Minkowskian case.
All the linear sets of generalized time-ordered functions we
consider are supposed to possess the standard translational
and spectral properties.
Supposing that we have linear sets $t^c$, $t^{1,c}$, $t^{2,c}$, of GTOF
having the factorization property (\ref{1.1}),
we consider all the GRF $r_\SS$
associated to $t^c$.
Their Fourier-Laplace transforms are all branches of a single
function $H$, holomorphic in a domain $D_X$ of
\begin{equation}
\LL_X =
\{k=(\{k\}_A,\ \{k\}_B,\ k_1) \in \bC^{d|X|}\ :\ 
k_1+ k_A +k_B =0\}
\label{h.1}\end{equation}
(Recall that for any $J\subset X$, $\{k\}_J$ stands for the set of
variables indexed by the elements of $J$, while $k_J= \sum_{j\in J}k_j$.)
We denote $\LL_X^{(r)} = \LL_X \cap \bR^{d|X|}$.
For each cell $\SS$, the Laplace transform of $r_\SS$ is the restriction 
of $H$ to the tube
\begin{equation}
\LL_X^{(r)} + i K_\SS =
\{k=p+iq \in \LL_X\ :\ \forall I\in \SS, \ \  q_I \in V_+\}.
\label{h.2}\end{equation}
The Fourier transform $\wt r_\SS$ of $r_\SS$ (with $\delta(p_X)$ removed)
is thus the boundary
value of $H$ from the tube (\ref{h.2}).

We denote
\begin{equation}
H(\{k\}_A,\ \{k\}_B,\ k_1) = h(\{k\}_A,\ \{k\}_B).
\label{h.3}\end{equation}
There also exist two functions
$H_1$ and $H_2$, the momentum-space analytic functions associated
to the GTOF $t^{1,c}$ and $t^{2,c}$ (appearing in (\ref{1.1})), 
depending on variables
labelled by $A\cup \{1\}$ and $B\cup \{1\}$,
respectively, such that 
\begin{equation}
H_1(\{k\}_A,\ -k_A) = h_1(\{k\}_A),\ \ \ 
H_2(\{k\}_B,\ -k_B) = h_2(\{k\}_B),
\label{h.4}\end{equation}
and we have
\begin{equation}
h(\{k\}_A,\ \{k\}_B) = h_1(\{k\}_A)\, h_2(\{k\}_B).
\label{h.5}\end{equation}
This can be rewritten as
\begin{equation}
H(\{k\}_A,\ \{k\}_B,\ k_1) =
H_1(\{k\}_A,\ k_1 + k_B)\, H_2(\{k\}_B,\ k_1 + k_A).
\label{h.6}\end{equation}
While this factorization follows from Corollary \ref{grffac},
it can also be understood without invoking 
Theorem \ref{algfac} or Corollary \ref{grffac}. Indeed the domain $D_X$  
contains the real open set 
$\RR = \{p \in \LL_X^{(r)}\ :\ p_I^2 < 0\ \ \forall I \in \PP_*(X)\}$,
and $H$ coincides there with the Fourier transform of $t_X^c$ (with
$\delta(p_X)$ removed). The latter factorizes as indicated in
(\ref{1.6}-\ref{1.10}), and this implies (\ref{h.6}) by analytic
continuation.

If $q$ is in the cone $K_\SS$, for every $I \subset A$ which belongs
to $\SS$, (resp. to $\SS'$), $q_I \in V_+$ (resp. $q_I \in V_-$).
This suffices to prescribe the ``sign'' of $q_J$ for every
proper subset $J$ of $A\cup\{1\}$, and we see that the point
$q= (\{q\}_A,\ q_1+q_B) \in \LL_{A\cup \{1\}}^{(r)} $ 
is in the cone $K_{\SA}$, where $\SA$ is precisely the cell 
associated to $A\cup\{1\}$ which is given by (\ref{3.3}).
Therefore the boundary value of $H_1(\{k\}_A,\ k_1 + k_B)$
from the tube $\LL_X^{(r)} +iK_S$ is the Fourier transform of $r_\SA$
(without $\delta$ function). Similarly the boundary value of 
$H_2(\{k\}_B,\ k_1 + k_A)$ from the same tube
is the Fourier transform of $r_\SB$. Thus, using (\ref{h.6}),
\begin{align}
&\wt r_S(\{p\}_A,\ \{p\}_B,\ p_1) =
\lim_{q\in K_\SS,\ q\rightarrow 0}H(\{p+iq\}_A,\ \{p+iq\}_B,\ p_1+iq_1)\cr
&=\wt r_\SA(\{p\}_A,\ p_1+p_B) \wt r_\SB(\{p\}_B,\ p_1+p_A)\ .
\label{h.7}\end{align}
Using (\ref{1.3}-\ref{1.10}) we recover the identity
\beq
r_\SS = r_\SA\,r_\SB
\label{h.7.1}\endq
obtained by algebraic arguments
in the preceding subsection.

\vskip 0.5 cm
We can now ask about
restricting $H$ (more precisely the restriction of $H$ to 
the tube $\LL_X^{(r)} + i K_\SS$) to the manifold $\{k\ :\ k_1 = 0\}$.

\vskip 0.25 cm
Let us assume first that $\SS = (A\cup B) \downarrow 1$. 
In this case we see immediately that if $q \in K_\SS$,
then $q_A$ and $q_B$ must both be in $V_-$ hence
$k_A + k_B$ can never vanish in the tube $\LL_X^{(r)}+iK_\SS$
and it is therefore not possible to restrict the restriction of
$H$ to this tube to $\{k\ :\ k_1 = 0\}$ or $\wt r_\SS$ to
$\{p\ :\ p_1 = 0\}$.

\vskip 0.25 cm
We now take $\SS = 1 \downarrow \TT$, 
$\SS' = 1 \uparrow \TT'$, where $\TT$ and $\TT'$ are two opposite cells 
relative to $A \cup B$. In this case $\DTA$ and $\UTPA$ are given in
(\ref{h.12}) and (\ref{h.13}), and similarly for $\DTB$ and $\UTPB$.

If $k = p+iq \in \LL_{A\cup B}^{(r)} + i K_\TT$, then
$(\{k\}_A,\ k_B)\in \LL_{A \cup 1} + i K_{\DTA}$
and 
$(\{k\}_B,\ k_A)\in \LL_{B \cup 1} + i K_{\DTB}$.
It is therefore possible to restrict the restriction of $H$ to
$\LL_X^{(r)} + i K_{1 \downarrow \TT}$ to the submanifold
$\{k_1 = 0\}$. We note that the actual reason is that
$A$ and $B$ are complementary in $A\cup B$, hence $k_A$ and
$k_B$ must have opposite ``signs'' in $\TT$.

Momentum space analyticity 
thus provides, in the Minkowskian case, a shorter and clearer account
of the factorization property, but the algebraic formalism is necessary
to deal with more general space-times, even relatively simple ones
such as de Sitter space-time.

\subsection{Factorization and supports}
\label{supports}

We return to the general (not necessarily Minkowskian) case.
Let again $X = A\cup B\cup\{1\}$, $A$, $B$ and $\{1\}$ disjoint,
$A \not= \emptyset$, $B \not= \emptyset$, and suppose given a
linear system of GTOF, indexed by
$X$, which has the factorization property (\ref{1.1}).
Let $\TT$ and $\TT'$ be opposite cells relative to $A\cup B$
and, as above, $\SS = 1 \downarrow \TT$, $\SS' = 1 \uparrow \TT'$.
According to Corollary \ref{grffac}, under our assumptions
(\ref{1.1}),
\begin{equation}
r_{1 \downarrow \TT}  = r_{\DTA}\,
r_{\DTB}\ ,
\label{s.1}\end{equation}
where the cells $\DTA$ (relative to $A\cup \{1\}$)
and $\DTB$ (relative to $B\cup \{1\}$)
are given by (\ref{h.12}) and (\ref{h.13}).
Suppose first that $A\in \TT$, and hence $B\in \TT'$.
Then $A \in \DTA$, hence 
$\{1\}$ belongs to the opposite cell $\UTPA$,
$B \in \UTPB$ hence $\{1\} \in \DTB$.
As a consequence, by (\ref{s.0.2}),
\begin{align}
&{\rm support}\ r_{\DTA}
\subset \bigcup_{j \in A}
\{\{x\}_{A\cup\{1\}}\ :\ x_1 \le x_j\}\ ,
\label{s.2}\\
&{\rm support}\ r_{\DTB}
\subset \bigcup_{k \in B}
\{\{x\}_{B\cup\{1\}}\ :\ x_k \le x_1\}\ ,
\label{s.2.1}\end{align}
\beq
{\rm support}\ r_{1\downarrow \TT} \subset
\bigcup_{j\in A}\bigcup_{k\in B}\{x\ :\ x_k\le x_1\le x_j\}\ .
\label{s.2.2}\endq
If $A\in \TT'$, and hence $B\in \TT$, the roles of $A$ and $B$ are exchanged,
and
\beq
{\rm support}\ r_{1\downarrow \TT} \subset
\bigcup_{j\in A}\bigcup_{k\in B}\{x\ :\ x_j\le x_1\le x_k\}\ .
\label{s.2.3}\endq
Thus, in all cases, $r_{1\downarrow \TT}$ vanishes unless $x_1$ 
remains in a finite union of double-cones (depending on the other variables). 
If the space-time $\XX$ is such that all double-cones are compact,
it is possible to integrate $r_{1\downarrow \TT}$ in the variable $x_1$,
obtaining a distribution (denoted $\wh r_\TT$) in the remaning variables. 

Recall that the relation $x_k \le x_j$ need not be an order
relation, and the above result would be valid even in
e.g. the Buchholz-Fredenhagen framework \cite{BF2} (in this case this relation
depends on the indices $j$ and $k$). 

\subsection{Additional properties when $\le$ is an order relation}
\label{ordrel}
The properties (\ref{s.0.1}) and
(\ref{s.0.2}) have much stronger consequences in the cases when
the relation $x_k \le x_j$ is an order relation, e.g. the Minkowski and
de Sitter spaces and the covering of the Anti-de Sitter space.

If $\le$ is an order relation,
and if $X$ and $Y$ are disjoint sets of indices, and
$\SS$ is a cell relative to $X$, then (see \cite{EGS2}) the support of
$r_{Y\updownarrow \SS}$ is contained in that of $r_\SS$, and
\begin{align}
{\rm support}\ r_{Y \downarrow \SS} &\subset
\{\{y\}_Y,\ \{x\}_X\ :\ 
\{x\}_X \in {\rm support}\ r_\SS\ ,\cr 
&\forall k \in Y\ \exists j \in X\ {\rm s.t.}\ 
y_k \le x_j \}\label{s.4}\\
{\rm support}\ r_{Y \uparrow \SS} &\subset
\{\{y\}_Y,\ \{x\}_X\ :\ 
\{x\}_X \in {\rm support}\ r_\SS\ ,\cr
&\forall k \in Y\ \exists j \in X\ {\rm s.t.}\ 
x_j \le y_k \}
\label{s.5}\end{align}
Furthermore if the variables indexed by $Y$ are kept fixed, the
set of the $r_{Y\downarrow \SS}$ (resp. $r_{Y\uparrow \SS}$) (as 
$\SS$ runs over all the cells associated to $X$) has all the linear
properties of a set of GRF associated to $X$. In particular the distributions
$\wh r_\SS$ obtained in the preceding subsection by integrating over $x_1$
have all the linear properties of a set of GRF 
associated to $X \setminus \{1\}$.

{\it From now on we shall restrict our attention to the cases when 
the relation $\le$ is an order relation.} 

\subsection{Integrating over several variables}

\begin{figure}[ht]
\begin{center}
\includegraphics[scale=0.5]{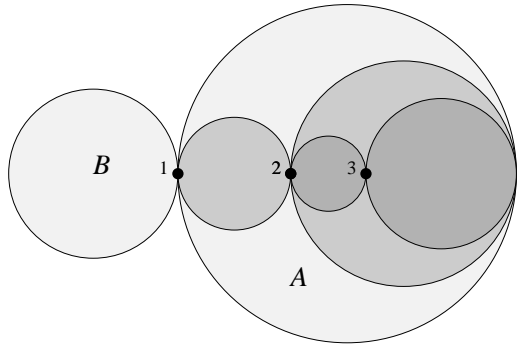}
\caption{Integration can be performed first on 3, then 2, then 1}
\label{recurf}
\end{center}
\end{figure}

The factorization formula (\ref{3.7}), applied to the case
$\SS = 1 \downarrow \TT$, $\SS' = 1 \uparrow \TT'$,
where $\TT$ and $\TT'$ are opposite cells relative to $A\cup B$,
(see eqs (\ref{h.10}) and (\ref{h.11})) gave the formulae
(\ref{h.12}) and (\ref{h.13}) for the cells
$\DTA$ and $\UTPA$.
We now wish to specialize this to the case when $2 \in A$, 
$A \setminus \{2\} = C \not= \emptyset$, $\VV$ and $\VV'$ are
opposite cells relative to $C\cup B$, and
\begin{equation}
\TT = 2 \downarrow \VV
= \{J \in \PP_*(A \cup B)\ :\ J = C\cup B\ {\rm or}\ 
J \cap (C\cup B) \in \VV\},
\label{v.1}\end{equation}
\begin{equation}
\TT' = 2 \uparrow \VV'
= \{J \in \PP_*(A \cup B)\ :\ J = \{2\}\ {\rm or}\ 
J \cap (C\cup B) \in \VV'\}.
\label{v.2}\end{equation}
Specializing (\ref{h.13}) to this case gives:
\begin{eqnarray}
\lefteqn{
\UUVPA =}\nonumber\\
&& \{J \in \PP_*(A \cup \{1\})\ :\ J = \{2\} \ {\rm or}\ 
J_1 = J\setminus \{2\}\ {\rm satisfies}\ :
\nonumber\\
&&(J_1 \subset C\ {\rm and}\ J_1 \in \VV')\ 
{\rm or}\ (1 \in J_1\ {\rm and}\ C\setminus J_1 \in \VV)\}
\nonumber\\
&&= 2 \uparrow \UVPC\ ,
\label{v.3}\end{eqnarray}
and hence
\begin{equation}
\DDVA =
2 \downarrow \DVC .
\label{v.4}\end{equation}
If all double-cones are bounded, this makes it possible  
to integrate recursively the expression
$r_{Y \updownarrow \TT}$ associated to a ``tree'' of the type symbolized
in Fig \ref{recurf}
over all 
variables labelled by $Y$, provided each $j \in Y$ is a
``splitting vertex'' for the ``tree''. We do not go into details
on this subject, but will focus on the special case provided
by actual tree graphs.

\section{Actual trees}
\label{actree}
The first part of this section does not require the assumption
that $\le$ is an order relation.
In this section we consider the linear system of
GTOF associated with a tree: this is a connected and simply connected
graph $Q$ whose vertices are labelled by a finite set of indices $X$
($|X| \ge 2$), which we take as $\{1,\ 2,\ldots,\ |X|\}$.
Each line $\ell$ of the tree has two distinct ends (also called extremities) 
in $X$, denoted
$b(\ell)$ and $e(\ell)$ with $b(\ell) < e(\ell)$. An extremity of $Q$
is a vertex which is the extremity of a single line. A vertex is
called internal if it is not an extremity of the tree, i.e. it is an end
to at least two distinct lines. A tree with more than one vertex
always has at least two extremities. If a line $\ell$ of the tree $Q$
is deleted, the resulting graph is the union of two disjoint tree graphs.
If their sets of vertices are denoted $Y$ and $Z$ we denote
$Q|Y$ and $Q|Z$ the corresponding trees.

For each line $\ell$ of $Q$, let $F_\ell$ 
be a two-point distribution having the PR property. Then
the product 
\beq
G = \prod_\ell F_\ell(x_{b(\ell)},\ x_{e(\ell)})
\label{tr.0}\endq
is well-defined and also has the PR property. This can be seen by induction
on the number of vertices $|X|$ of $Q$ by the same argument 
as in subsect. \ref{fac1}.

The time-ordered function associated to the tree is
\begin{equation}
t_X(x_1,\ldots,\ x_{|X|}) = \prod_\ell t_\ell(x_{b(\ell)},\ x_{e(\ell)})
\label{tr.1}\end{equation}
where the product extends over all the lines of the tree $Q$,
and $t_\ell(u,\ v)$ is the time-ordered two-point function of some free
or generalized free field $\phi_\ell$. 
Let 
$w_{\ell+}(u,\ v) = (\Omega,\ \phi_\ell(u)\phi_\ell(v)\Omega) = w_{\ell-}(v,\ u)$.
We assume that all the $t_\ell$ and $w_{\ell\pm}$ have the PR property.
If $a = \{J_1,\ldots,J_\nu\}$ is a proper sequence in $\PP(X)$,
\begin{align}
t_a(x_1,\ldots,\ x_{|X|}) &= 
t_{J_1,\ldots,J_\nu}(x_1,\ldots,\ x_{|X|}) 
= \prod_\ell F_\ell(x_{b(\ell)},\ x_{e(\ell)})\ ,\cr
F_\ell(x_{b(\ell)},\ x_{e(\ell)}) &= t_\ell(x_{b(\ell)},\ x_{e(\ell)})\ \ 
\hbox{if $b(\ell)$ and $e(\ell)$ belong to the same $J_k$},\cr
F_\ell(x_{b(\ell)},\ x_{e(\ell)}) &= w_{\ell+}(x_{b(\ell)},\ x_{e(\ell)})\ \ 
\hbox{if $b(\ell) \in J_r$, $e(\ell) \in J_k$ with $r<k$},\cr
F_\ell(x_{b(\ell)},\ x_{e(\ell)}) &= w_{\ell-}(x_{b(\ell)},\ x_{e(\ell)})\ \ 
\hbox{if $b(\ell) \in J_r$, $e(\ell) \in J_k$ with $r>k$}\ .
\label{tr.2}\end{align}
Again the product (\ref{tr.2}) is well-defined because each of the 
two-point functions
involved has the PR property and beecause we are dealing with a tree graph.
According to preceding definitions
\begin{align}
&r_{e(\ell) \uparrow b(\ell)}(x_{b(\ell)},\ x_{e(\ell)}) = 
t_\ell(x_{b(\ell)},\ x_{e(\ell)})-w_{\ell+}(x_{b(\ell)},\ x_{e(\ell)}),\cr
&r_{e(\ell) \downarrow b(\ell)}(x_{b(\ell)},\ x_{e(\ell)}) = 
t_\ell(x_{b(\ell)},\ x_{e(\ell)})-w_{\ell-}(x_{b(\ell)},\ x_{e(\ell)})\ .
\label{tr.3}\end{align}
The condition that the $t_a$'s have the causal factorization property
is that:
\beq
{\rm support}\ r_{e(\ell) \uparrow b(\ell)} \subset 
\{(x,\ y)\in \XX^2\ :\ x \le y\},\ \ {\rm support}\ r_{e(\ell) \downarrow b(\ell)}
\subset \{(x,\ y)\in \XX^2\ :\ x \ge y\}\ .
\label{tr.3.1}\endq

\begin{lemma}
\label{treefac}
Let $Q$ be a tree with vertices indexed by $X$ as above, and
$\SS$ be a cell associated to $X$, opposite to $\SS'$.
For each line $\ell$ of $Q$ with ends $b(\ell)$ and $e(\ell) > b(\ell)$,
let $B(\ell)$ and $E(\ell)$ be the sets of vertices of the two
subtrees of $Q$ 
obtained by severing the line $\ell$ and such that
$b(\ell) \in B(\ell)$, $e(\ell) \in E(\ell)$.
The GRF $r_\SS$ associated to $Q$  (i.e. obtained from the GTOF
defined in (\ref{tr.2}))
is a product over the lines of $Q$
\begin{equation}
r_\SS = \prod_{\ell} r_{b(\ell) \updownarrow e(\ell)}.
\label{tr.4}\end{equation}
(i) If $e(\ell)$ is an extremity of the tree $Q$ 
(i.e. if $E(\ell) = \{e(\ell)\}$) the factor contributed by the line
$\ell$ is $r_{e(\ell) \uparrow b(\ell)}$ if
$\{e(\ell)\} \in \SS$ while it is $r_{e(\ell) \downarrow b(\ell)}$ 
if $\{e(\ell)\} \in \SS'$.\HB
(ii) More generally, the factor contributed by the line $\ell$ is
$r_{b(\ell) \uparrow e(\ell)}$ if $B(\ell) \in \SS$,
and $r_{b(\ell) \downarrow e(\ell)}$ if $B(\ell) \in \SS'$.
\end{lemma}

\begin{figure}[ht]
\begin{center}
\includegraphics[scale=0.8]{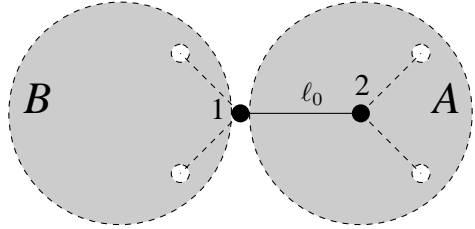}
\vglue -4 cm
\setlength{\unitlength}{0.8 cm}
\begin{picture}(16,5)
\put(9.0, 2.75){$\ell_0$}
\end{picture}
\caption{\label{spltree}The tree $Q$}
\end{center}
\end{figure}

\noindent {\bf Proof.}
Repeated application of the factorization theorem 
(Corollary \ref{grffac})
shows that the GRF associated to $Q$ also factorize as products
of two-point GRF.
We suppose $|X| > 2$. Let $\SS$ be a cell
associated to $X$. We may suppose without loss of generality
that some line $\ell_0$ of the tree $Q$ has 1 and 2 as ends (i.e.
$b(\ell_0) = 1$, $e(\ell_0)= 2$), and that 1 is not an extremity
of $Q$. Severing $\ell_0$ produces two disjoint subtrees
of $Q$, whose sets of vertices are denoted $B\cup\{1\}$ and
$A$, respectively, with $A\cup B\cup\{1\} = X$,
$A$, $B$, and $\{1\}$ disjoint, $B \not= \emptyset$, $2 \in A$.
By Corollary \ref{grffac}, the GRF $r_\SS$ for $Q$ factorizes into
\begin{equation}
r_{\SS} = r_{\SA}\,r_{\SB},
\label{tr.5}\end{equation}
where $r_{\SA}$ is the GRF associated to the tree
$Q|A\cup\{1\}$ and the cell $\SA$, and
$r_{\SB}$ is the GRF associated to the tree
$Q|B\cup\{1\}$ and the cell $\SB$.
If $A \in \SS$ then $A \in \SA$ hence $\{1\} \in \SPA$,
while if $A \in \SS'$, then $\{1\} \in \SA$
(see Figure \ref{spltree}).

Assume first that 2 is an extremity of $Q$, that is,
$A=\{2\}$. Then if $\{2\} \in \SS$, we find
${\SA} = 2\uparrow 1$, while if if $\{2\} \in \SS'$, we find
${\SA} = 1\uparrow 2 = 2\downarrow 1$.
This proves part(i) of the lemma.

We now return to the general case (2 not necessarily an
extremity of $Q$) and note that $1$ is now an extremity of
$Q|A\cup\{1\}$. By the above, the contribution of the
line $\ell_0$ to $r_\SA$, hence to the $r_\SS$ associated to $Q$,
is $r_{1\uparrow 2}$ if $A \in \SS'$, and $r_{1\downarrow 2}$
if $A \in \SS$. This proves part(ii).\ \endprf

Another way to express the content of Lemma \ref{treefac} is
to represent the GRF $r_\SS$ associated to the graph $Q$ and the 
cell $\SS$ by orienting the lines of $Q$: the line $\ell$
will be oriented (e.g. by drawing an arrow) 
from $b(\ell)$ to $e(\ell)$ if $E(\ell) \in \SS$,
and in the opposite direction if $B(\ell) \in \SS$.
After this, if a line joins the vertices $j$ and $k$ and its arrow
points towards $k$, this line contributes $r_{k\uparrow j}$ and
the support of $r_\SS$ is contained in $\{x\ :\ x_j \le x_k\}$,
i.e. the arrow points to the future. 
Still supposing that 1 is an internal vertex, we now assume
$\SS = 1 \downarrow \TT$ where $\TT$ is a cell relative to
$X\setminus \{1\}$. Let $\ell_0, \ldots,\ \ell_p$ be the lines
having 1 as an extremity. Since 1 is internal, $p \ge 1$. We assume that
(in accordance with the notations of this subsection)
$b(\ell_j) =1$ for $0 \le j \le p$, and the sets
$E(\ell_j)$ are defined as in Lemma \ref{treefac}.
These sets are disjoint and non-empty and 
$\bigcup_0^p E(\ell_j) = X\setminus \{1\}$.
For any $j$, $\ell_j$ is oriented away from 1 if $E(\ell_j) \in \SS$,
and towards 1 if $E(\ell_j) \in \SS'$. By the definition of 
$1 \downarrow \TT$, $E(\ell_j) \in \SS \Leftrightarrow E(\ell_j) \in \TT$
and $E(\ell_j) \in \SS' \Leftrightarrow E(\ell_j) \in \TT'$.
At least one
$E(\ell_j)$ belongs to $\TT$, and at least another one, $E(\ell_k)$,
belongs to $\TT'$. Hence at least one of the lines $\ell_j$
is directed away from 1, giving a contribution 
$r_{1\downarrow e(\ell_j)}$, and another one $\ell_k$ is
directed towards 1, giving a contribution $r_{1\uparrow e(\ell_k)}$.
Thus the support of $r_\SS$ is contained in
$\{\{x\}_X\ :\ x_{e(\ell_k)} \le x_1 \le x_{e(\ell_j)}\}$. 
As mentioned before, this conclusion does not hold for general $\SS$,
and it is false for $\SS = (X\setminus \{1\}) \updownarrow 1$.

We again assume, from now on, that $\le$ is an order relation.
Let us suppose that $X = Y\cup Z$, $Y\cap Z = \emptyset$, 
$Y\not= \emptyset$ (say $1\in Y$), $Z\not= \emptyset$, all vertices labelled by
$Y$ being internal (some of the $Z$-vertices may also be internal,
but $Z$ contains all the extremities of $Q$). Let 
$\SS = Y\downarrow \VV$, where $\VV$ is a cell associated to $Z$.
We assign a direction (arrow) to each line as described above.
For each $j\in Y$ we have 
$\SS = j\downarrow (Y\setminus \{j\})\downarrow \VV)$. Hence we
may apply to the index $j$ the same argument that was applied above
to the case $j=1$. Therefore there is at least one line having $j$ as
extremity and directed towards $j$ and another line directed away
from $j$. If the other vertex $k$ of this line is in $Y$, we can continue
this line by another one directed away from $k$ and so on until
we reach a $Z$-vertex. We may proceed in the same manner downwards from
$j$, and (because $\le$ is an order relation) there are two indices
$a\in Z$ and $b\in Z$ such that the support of $r_\SS$ is contained in
$\{\{x\}_X\ :\ x_a \le x_j \le x_b\}$. Thus the variable indexed
by any $Y$-vertex is confined to a double-cone with $Z$-vertices as
its vertices. If the double-cones are all compact in $\XX$ (as it
is the case in the Minkowski and de Sitter space-times) one can
integrate over all variables indexed by $Y$, and the result has all
the properties of a GRF associated to $Z$ and the cell $\VV$.

\section{Perturbation theory. Conclusion}
\label{perth}
In perturbation theory, solving the UV problem for a certain theory
provides time-ordered products for Wick monomials
in a finite number of free (or generalized free) fields. A connected Feynman
graph\footnote{We only consider graphs in which each line has
two distinct ends}
$G$ with vertices indexed by $X = \{1,\ldots,\ n\}$
represents one of the contributions to the (truncated) vacuum 
expectation value of a time-ordered product of $|X|$ Wick monomials
of free (or generalized free) fields, e.g.
\beq
(\Omega,\ T(\psi_1(x_1)\ldots \psi_n(x_n))\Omega)_c\,.
\label{p.10}\endq
However we may assign a different
free field $\phi_\ell$ to each line $\ell$ of $G$ (with
$[\phi_\ell,\ \phi_{\ell'}] = 0$ if $\ell \not= \ell'$) and redefine
\beq
\psi_j(x_j) = \prod_{\hbox{$\ell$ : $j$ is an end of $\ell$}}\phi_\ell(x_j)\ .
\label{p.20}\endq
(We decorate with derivatives and additional indices as needed.)
With this definition, $G$ 
represents the only contribution to (\ref{p.10}).
The construction of time-ordered products also provides a
time-ordered product for any subset of the fields $\psi_j$, so that
$G$ is associated to a full linear system of GTOF:
\beq
t^c_{J_1,\ldots,J_\nu} =
(\Omega,\ T(J_1)\ldots T(J_\nu)\Omega)_c\ ,
\label{p.21}\endq
\beq
J_1 \cup \ldots\ J_\nu = X,\ \ \ 
J_j \cap J_k = \emptyset\ {\rm if}\ j \not= k\ ,
\label{p.22}\endq
\beq
T(J_r) = T(\prod_{j\in J_r}\psi_j(x_j))\ .
\label{p.23}\endq
From this one can obtain the contribution of the graph to all possible 
generalized retarded functions for these Wick monomials.
A simple example of this situation has been seen in the case when $G$ is a tree
graph in Sect. \ref{actree}. (Of course we are adhering here to the most
simplistic view of perturbation theory, and the reader might enjoy, by
contrast, the point of view of \cite{S3} and references therein).

The contribution of such a graph to the GTOF
for the interacting fields is obtained by distinguishing
two groups of variables indexed by $Y$ and $Z$ (where
$Y \cup Z = X$, $Y \cap Z = \emptyset$). The $Y$ variables
correspond to the interactions and the associated Wick
monomials are assumed to have degree $>1$ (all the $Y$ vertices are
internal). One should then
perform the integration denoted symbolically
\begin{equation}
\wh t^c_{J_1,\ldots,J_\nu} = \int_{\XX^Y}\, dY\ 
t^c_{Y \downarrow \{J_1,\ldots,J_\nu\}}
\label{p.3}\end{equation}
Here $J_1\cup \ldots \cup J_\nu = Z$, and hatted quantities refer to 
interacting fields.
In particular, if $\TT$ is a cell relative to $Z$,
\begin{equation}
\wh r_\TT = \int_{\XX^Y}\, dY\ r_{Y \downarrow \TT}
\label{p.4}\end{equation}
The weak adiabatic limit problem consists in making sense 
of the $Y$ integration.
In the Minkowskian case it is easy (see e.g. \cite{EG}) 
when there are no zero masses.

What we have seen in Sect \ref{actree} is that, for tree graphs, 
if double-cones are bounded, it is always possible to
obtain the contribution of the graph to the GRF of the interacting fields.
Obtaining from this the GTOF (in particular the
Wightman functions) requires a splitting of multiple commutators:
for example for a 2-point function in the Minkowski case, given 
the commutator function $\wh c(x_2- x_1)$,
find $\wh w_+(x_2-x_1)$ and $\wh w_-(x_2-x_1)$, respectively
holomorphic in the future and past tubes, such that
$\wh w_+ - \wh w_- = \wh c$. In this case it is easy to find solutions,
but this is not known for general $n$-point functions.

\section*{Acknowledgements}
I thank Jacques Bros and Ugo Moschella for very helpful discussions.

\appendix
\section{
Appendix. 
Some simple examples}
\label{app1}
Let $X = \{1,\ 2,\ 3,\ 4\}$. All cells relative to $X$ are
Steinmann monomials \cite{S1,S2}. They are given by
\begin{equation}
\begin{array}{ccc}
a_j = k\uparrow m\uparrow n\uparrow j & &
r_j = k\downarrow m\downarrow n\downarrow j \\
a_{jk} = k\downarrow m\uparrow n\uparrow j & &
r_{jk} = k\uparrow m\downarrow n\downarrow j
\end{array}
\label{ap.1}\end{equation}
where $(j,\ k,\ m,\ n)$ is any permutation of $(1,\ 2,\ 3,\ 4)$.
(More precisely $a_j$, $a_{jk}$, $r_j$, $r_{jk}$ denote the corresponding
GRF, while $j\updownarrow k\updownarrow m\updownarrow  n$
denote cells. This will produce no confusion here.)
Figure \ref{stplanet} illustrates the 4-point geometrical cells.
\setlength{\unitlength}{0.8 cm}
\begin{figure}[h!]
\begin{center}
\def\spot{{\kern -0.2em\lower.55ex\hbox{$\bullet$}}}

\begin{picture}(16,10)
\put(3.40, 2.30){$a_{14}$}
\put(3.40,6.00){$a_4$}
\put(3.40,4.00){$a_{41}$}
\put(5.00,3.00){$r_{32}$}
\put(1.50,3.00){$r_{23}$}
\put(0.50,4.30){$r_2$}
\put(6.50,4.30){$r_3$}
\put(0.00,6.00){$r_{21}$}
\put(6.50,6.00){$r_{31}$}
\put(1.50,7.00){$a_{43}$}
\put(5.30,7.00){$a_{42}$}
\put(0.00,7.70){$a_{34}$}
\put(6.40,7.70){$a_{24}$}
\put(1.50,8.90){$r_{12}$}
\put(5.00,8.90){$r_{13}$}
\put(3.40,9.30){$r_1$}
\put(12.40,9.70){$r_{14}$}
\put(12.40,6.00){$r_4$}
\put(12.40,8.00){$r_{41}$}
\put(11.00,9.20){$a_{23}$}
\put(13.85,9.20){$a_{32}$}
\put(15.20,7.60){$a_3$}
\put(9.50,7.60){$a_2$}
\put(15.70,6.00){$a_{31}$}
\put(9.00,6.00){$a_{21}$}
\put(14.10,4.60){$r_{42}$}
\put(10.40,4.60){$r_{43}$}
\put(15.50,4.25){$r_{24}$}
\put(9.00,4.25){$r_{34}$}
\put(14.00,3.10){$a_{13}$}
\put(10.60,3.10){$a_{12}$}
\put(12.40,2.70){$a_1$}
\includegraphics[scale=0.8]{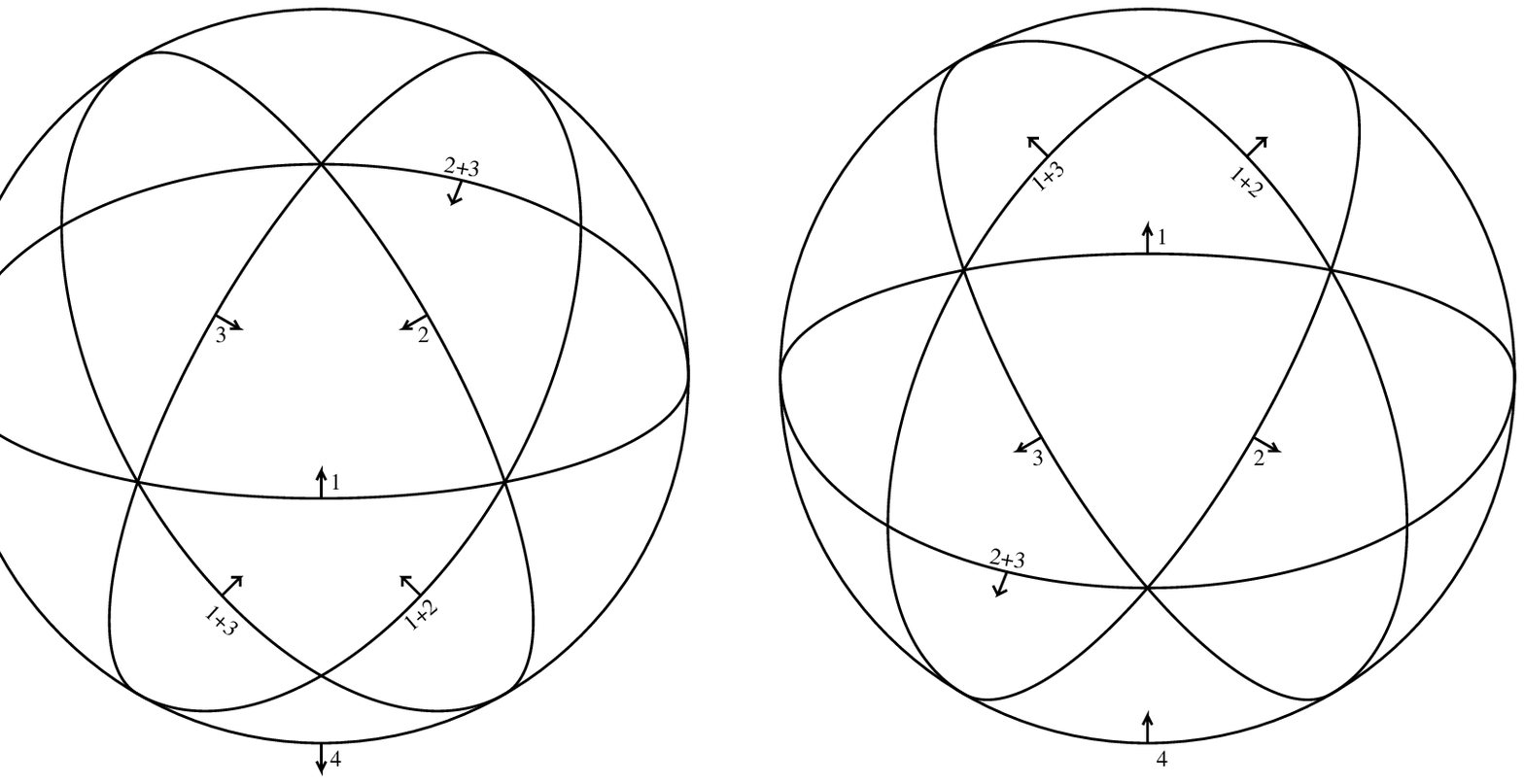}
\end{picture}
\vglue -1cm
\caption{\label{stplanet}
\lsmall
The Steinmann Planet.
The unit sphere in $\{(s_1,\ldots,s_4) \in \bR^4\ :\ \sum_{j=1}^4 s_j =0\}$
intersects the planes $\{s_J =0\}$ along great circles. The geometrical cells
are the polyhedral open cones which are the connected components
of the complement of the union of these planes.
}
\end{center}
\end{figure}

Table \ref{table1} shows how this applies to the 4-point GRF
associated with the very simple tree graph
\begin{equation}
\hbox{Graph 1 :}\ \ \ \bu{1}\lineto\udbu{4}{3}\lineto\bu{2}
\label{ap.2}\end{equation}
By virtue of the symmetric role played by the indices 1, 2, and 3, 
and the rules of the arrow calculus, all the cells not mentioned
in the table can be obtained from those mentioned by
a permutation of the indices 1, 2, 3, or by reversing all arrows, or both.

\begin{table}[h!]
\begin{center}
\def\stretcher{{\vrule height 0.5cm width 0cm depth 0.9cm}}
\begin{tabular}{|c|c|c|}
\hline
Cell & Support of cell & Factorization \\
\hline
$4\downarrow 1\downarrow 2\downarrow 3$ ($r_3$) &
$x_j \le x_3$ for all $j= 4,\ 1,\ 2$ &
\stretcher $\bu{1}\arto\downbu{4}{3}\arfr\bu{2}$\\
\hline
$4\downarrow 1\uparrow 2\uparrow 3$ ($a_{34}$) &
$x_4 \le x_j$ for $j=1$ or 2 or 3, $x_1 \ge x_3$, $x_2 \ge x_3$ &
\stretcher $\bu{1}\arfr\upbu{4}{3}\arto\bu{2}$\\
\hline
$1\uparrow 4\downarrow 2\downarrow 3$ ($r_{31}$) &
$x_1 \ge x_j$ for $j=4$ or 2 or 3, $x_4 \le x_3$, $x_2 \le x_3$ &
\stretcher $\bu{1}\arfr\downbu{4}{3}\arfr\bu{2}$\\
\hline
$1\uparrow 2\uparrow 3\uparrow 4$ ($a_4$) &
$x_j \ge x_4$ for all $j= 1,\ 2,\ 3$ &
\stretcher $\bu{1}\arfr\downbu{4}{3}\arto\bu{2}$\\
\hline
$1\downarrow 2\uparrow 3\uparrow 4$ ($a_{41}$) &
$x_1 \le x_j$ for $j=4$ or 2 or 3, $x_2 \ge x_4$, $x_3 \ge x_4$ &
\stretcher $\bu{1}\arto\downbu{4}{3}\arto\bu{2}$\\
\hline
\end{tabular}
\caption{\label{table1}Factorization table for the first example}
\end{center}
\end{table}

We now consider $Z = Y\cup X$, $X = \{1,\ 2,\ 3,\ 4\}$, $Y = \{0\}$
or $Y = \{0,\ 5\}$ and the cells $0\downarrow \SS$ and 
$0\downarrow 5\downarrow\SS$ respectively associated with the tree graphs:
\beq
\hbox{Graph 2\ :}\ \ \ 
{\vrule height 1 cm width 0cm depth 1 cm}
\bu{1}\lineto\colbu{4}{0}{3}\lineto\bu{2} \qquad \qquad\qquad 
\hbox{Graph 3\ :}\ \ \ 
\colbu{1}{0}{2} \lineto \colbu{3}{5}{4}\ .
\label{ap.3}\endq
Here $\SS$ is a cell associated with $X$. Table \ref{table2}
shows the factorizations
and the resulting supports for a sample of the cells $\SS$

\begin{table}[h!]
\begin{center}
\def\stretcher{{\vrule height 1 cm width 0cm depth 1 cm}}
\begin{tabular}{|c|c|c|}
\hline
$\SS$ & $0\downarrow \SS$ for Graph 2 
& $0\downarrow 5\downarrow\SS$ for Graph 3\\
\hline
$4\downarrow 1\downarrow 2\downarrow 3$ ($r_3$) &
\stretcher $\bu{1} \arto \coluubu{3}{0}{4}\arfr\bu{2}$ &
$\coldubu{1}{0}{2} \arto \coluubu{3}{5}{4}$\\
\hline
$1 \uparrow 2 \uparrow 3 \uparrow 4$ ($a_4$) &
\stretcher $\bu{1} \arfr \coluubu{3}{0}{4} \arto \bu{2}$ &
$\coludbu{1}{0}{2} \arfr \coluubu{3}{5}{4}$\\
\hline
$1 \downarrow 2 \uparrow 3 \uparrow 4$ ($a_{41}$) &
\stretcher $\bu{1} \arto \coluubu{3}{0}{4} \arto \bu{2}$ &
$\colddbu{1}{0}{2} \arfr \coluubu{3}{5}{4}$\\
\hline
$1\uparrow 4\downarrow 2\downarrow 3$ ($r_{31}$) &
\stretcher $\bu{1} \arfr \coluubu{3}{0}{4} \arfr \bu{2}$ &
$\coluubu{1}{0}{2} \arto \coluubu{3}{5}{4}$\\
\hline
\end{tabular}
\caption{\label{table2}Factorization table for Graphs 2 and 3}
\end{center}
\end{table}

In particular the four-point generalized retarded functions for 
a scalar field with an interaction density $\LL(x) = :\phi^4(x):$
are given, in the first order of perturbation theory,
by
\beq
(\Omega,\ \wh R_\SS(\phi(x_1),\ldots,\ \phi(x_4))\,\Omega) =
\int_{\XX} (\Omega,\ 
R_{0 \downarrow \SS}(\LL(x_0),\ \phi(x_1),\ldots,\ \phi(x_4))\,\Omega)\,
dx_0\ ,
\label{ap.4}\end{equation}
and the integral is convergent by table \ref{table2}.

\newpage


\end{document}